\documentclass[conference]{IEEEtran}
\IEEEoverridecommandlockouts

\usepackage{cite}
\usepackage[a4paper, total={184mm,239mm}]{geometry}
\def\BibTeX{{\rm B\kern-.05em{\sc i\kern-.025em b}\kern-.08em
    T\kern-.1667em\lower.7ex\hbox{E}\kern-.125emX}}

\usepackage{amsmath}

\usepackage{amssymb,amsfonts}
\usepackage{algorithmic}
\usepackage{graphicx}
\usepackage{textcomp}
\usepackage{xcolor}
\def\BibTeX{{\rm B\kern-.05em{\sc i\kern-.025em b}\kern-.08em
    T\kern-.1667em\lower.7ex\hbox{E}\kern-.125emX}}
\usepackage[utf8]{inputenc}
\usepackage{booktabs}
\usepackage{graphicx}
\usepackage{subfigure}
\usepackage{units}
\usepackage{accents}
\usepackage{blindtext}
\usepackage{multirow}
\usepackage{xcolor,xspace,amsbsy}
\usepackage{amsmath}
\usepackage{paralist}
\usepackage{mdframed}
\usepackage{pict2e}
\usepackage{listings}
\usepackage{color}
\usepackage{threeparttable}
\usepackage{array}
\usepackage{booktabs}
\usepackage{pifont}
\usepackage{balance}
\usepackage{url}
\usepackage{tabularx}
\usepackage{caption}
\usepackage{lipsum}
\usepackage{makecell}
\usepackage[most]{tcolorbox}
\usepackage{boldline} 
\usepackage{tikz}
\usepackage{bm,utfsym}
\usepackage{makecell}%To keep spacing of text in tables
\usepackage{tabularray}
\usepackage{tablefootnote}
\usepackage{colortbl}
\usepackage{soul}
\usepackage{cellspace}
\usepackage{comment}
\usepackage{subcaption}
\usepackage{float}
\usepackage{hhline}
\usepackage{hyperref}

\usepackage[symbol]{footmisc}

\usepackage[ruled]{algorithm2e}
\usepackage{algorithmic}
\LinesNumbered

\usepackage{url}

\usepackage[
    n,
    advantage,
    operators,
    sets,
    adversary,
    landau,
    probability,
    notions,
    logic,
    ff,
    mm,
    primitives,
    events,
    complexity,
    asymptotics,
    keys]{cryptocode}

\usetikzlibrary{arrows, patterns, automata}

\definecolor{burgundy}{RGB}{144,0,32}
\definecolor{myblue}{rgb}{0.0, 0.2, 0.8}
\definecolor{mygreen}{rgb}{0,0.6,0}
\definecolor{mygray}{rgb}{0.5,0.5,0.5}
\definecolor{mymauve}{rgb}{0.58,0,0.82}
\definecolor{bluegray}{rgb}{0.4, 0.6, 0.8}
\definecolor{azure}{rgb}{0.0, 0.5, 1.0}
\definecolor{darkcandyapplered}{rgb}{0.64, 0.0, 0.0}

\newcommand{\casu}{{{\ensuremath{\mathsf CASU}}}\xspace}
\newcommand{\cfi}{{{\ensuremath{\mathsf CFI}}}\xspace}
\newcommand{\cfa}{{{\ensuremath{\mathsf CFA}}}\xspace}
\newcommand{\dfa}{{{\ensuremath{\mathsf DFA}}}\xspace}
\newcommand{\cacfi}{{{\sf EILID}}\xspace}
\newcommand{\cacfilong}{{{ {\underline E}xecution {\underline I}ntegrity for {\underline L}ow-end 
{\underline I}oT {\underline D}evices}}\xspace}
\newcommand{\cacfiinst}{\ensuremath{\mathsf{\it \cacfi_{inst}}}\xspace}
\newcommand{\cacfisw}{\ensuremath{\mathsf{\it \cacfi_{sw}}}\xspace}
\newcommand{\cacfihw}{\ensuremath{\mathsf{\it \cacfi_{hw}}}\xspace}

\newcommand{\rot}{\texttt{RoT}\xspace}

\newcommand{\prv}{{\ensuremath{\sf{\mathcal Prv}}}\xspace}
\newcommand{\vrf}{{\ensuremath{\sf{\mathcal Vrf}}}\xspace}
\newcommand{\ra}{{\ensuremath{\sf{\mathcal RA}}}\xspace}
\newcommand{\sadv}{{\ensuremath{\sf{\mathcal Adv}}}\xspace}

\newcommand{\propa}{{\bf{P1}}\xspace}
\newcommand{\propb}{{\bf{P2}}\xspace}
\newcommand{\propc}{{\bf{P3}}\xspace}

\mathchardef\mhyphen="2D

\begin{document}

\title{\cacfi: \cacfilong}

\author{\IEEEauthorblockN{Sashidhar Jakkamsetti}
\IEEEauthorblockA{
\textit{Robert Bosch LLC}\\
sashidhar.jakkamsetti@us.bosch.com}
\and
\IEEEauthorblockN{Youngil Kim}
\IEEEauthorblockA{
\textit{UC Irvine}\\
youngik2@uci.edu}
\and
\IEEEauthorblockN{Andrew Searles}
\IEEEauthorblockA{
\textit{UC Irvine}\\
searlesa@uci.edu}
\and
\IEEEauthorblockN{Gene Tsudik}
\IEEEauthorblockA{
\textit{UC Irvine}\\
gene.tsudik@uci.edu}
}

\maketitle

\begin{abstract}
Prior research yielded many techniques to mitigate software 
compromise for low-end Internet of Things (IoT) devices. Some of them
detect software modifications via remote attestation and similar services, while others 
preventatively ensure software (static) integrity. However, achieving run-time (dynamic) 
security, e.g., control-flow integrity (\cfi), remains a challenge.

Control-flow attestation (\cfa) is one approach that
minimizes the burden on devices. However, \cfa is not a real-time countermeasure
against run-time attacks since it requires communication with a verifying entity.
This poses significant risks if safety- or time-critical tasks have memory vulnerabilities.

To address this issue, we construct \cacfi\ -- a hybrid architecture that ensures software
execution integrity by actively monitoring control-flow violations on low-end devices. 
\cacfi is built atop \casu \cite{casu-brief}, a prevention-based (i.e., active) hybrid Root-of-Trust 
(\rot) that guarantees software immutability.  \cacfi achieves fine-grained backward-edge 
and function-level forward-edge \cfi via semi-automatic code instrumentation and a secure shadow stack.
\end{abstract}

\section{Introduction}\label{sec:intro}
A recent report \cite{num_of_iot_devs} states that there are now over $15$ billion IoT devices 
worldwide, and this number is predicted to double by 2030.  These devices are deployed in a wide range 
of settings, including homes, offices, farming, factories, public venues, and vehicles.
They also often collect sensitive information and/or perform safety-critical tasks. Unlike higher-end 
computing devices (e.g., laptops, tablets, and smartphones), IoT devices generally lack robust 
security features due to cost, size, energy, and performance constraints, making them 
attractive attack targets. Recent surveys demonstrate that IoT devices are subject to numerous attacks
through physical, network, software, and cryptographic vulnerabilities \cite{security_attacks_iot-brief,taxonomy_of_attacks-brief,meneghello2019iot-brief,ronen2016extended-brief,sikder2021survey-brief}.

To that end, various research proposed Roots-of-Trust ({\rot}s) for low-end devices 
\cite{pistis-brief,sancus-brief,vrasedp-brief,rata-brief,ida2024-brief,pure-brief,asokan2018assured-brief,surminski2021realswatt-brief,lazarus-effect-brief,garota-brief,awdt-dominance-brief,jin2022understanding-brief,alasmary2019analyzing-brief,vijayakanthan2023swmat-brief}. 
One popular technique underlying these proposals is Remote Attestation (\ra) 
\cite{pistis-brief,vrasedp-brief,sancus-brief,rata-brief,ida2024-brief,pure-brief,asokan2018assured-brief}, 
a well-established security service that {\bf detects} malware presence on an untrusted remote device. 
\ra requires a device (prover or \prv) to securely interact with a remote trusted party 
(verifier or \vrf), conveying the current software state of the former to the latter.
An alternative approach, exemplified by \casu\cite{casu-brief}, {\bf prevents} all software modifications 
except secure updates. However, all these detection- and prevention-based techniques focus on 
{\bf static} software integrity and do not provide any protection against, or detection of, run-time attacks. 

Since run-time attacks do not modify the software code itself, conventional \ra schemes can not handle them.
To mitigate such attacks, many Control-Flow Attestation (\cfa) 
\cite{cflat-brief,tinycfa-brief,caulfield2023acfa-brief,lofat-brief,ammar2024bridging-brief,yadav2023whole-brief,zhang2021recfa-brief,neto2023isc-brief,atrium-brief,debes2023zekra-brief,wang2023ari-brief} 
and Data-Flow Attestation (\dfa) \cite{litehax-brief,dialed-brief,abera2019diat-brief,kuang2020ra-brief}
schemes have been proposed. They allow \vrf to check execution integrity of remote \prv-s at 
run-time. However, merely detecting run-time attacks is insufficient, particularly in safety-critical 
environments where real-time response and recovery are imperative. Furthermore, practicality of \cfa itself
is questionable since it requires storing -- and later transferring -- potentially voluminous logs.

Another line of research \cite{davi2015hafix-brief, christoulakis2016hcfi-brief, de2019fixer-brief,
zhou2020silhouette-brief, nyman2017cfi-brief, sullivan2016strategy-brief, almakhdhub2020mu-brief, das2016fine-brief, 
liljestrand2019pac-brief, gollapudi2023control-brief,park2022bratter-brief,yoo2022kernel-brief,
murray2013sel4-brief,de2017sofia-brief,walls2019control-brief,fu2022fh-brief,peng2023cefi-brief}
produced various Control-Flow Integrity (\cfi) techniques for embedded systems.
However, all such efforts target higher-end devices with ample computing resources 
(e.g., multiple cores and many MBs of memory) and advanced security features, such as Memory 
Protection Units (MPUs), Memory Management Units (MMUs), or Trusted Execution Environments (TEEs), 
making them unsuitable for low-end devices.

To bridge this gap, this work constructs a real-time countermeasure, \cacfi, against control-flow attacks 
on low-end devices. \cacfi uses \casu as a foundational platform to offer user software immutability. 
It extends \casu by implementing \cfi monitor that ensures fine-grained backward-edge and function-level 
forward-edge execution integrity. \cacfi has three components:
(1) {\em Code instrumenter} -- inserts additional instructions into device software at 
compile-time to jump to \cfi monitor, 
(2) {\em CFI monitor} -- trusted (minimal) software that maintains a secure shadow stack
for validating crucial control-flow metadata, and
(3) {\em Secure hardware} -- derived from \casu, it detects control-flow violation and triggers a reset.

Contributions of this work are twofold:

  \begin{compactitem}
  % \noindent $\sbullet[.75]$ 
  \item To the best of our knowledge at the time of this writing, \cacfi is the first \rot architecture enforcing 
  \cfi on low-end devices, given that relevant prior work either focused on detecting such attacks 
  \cite{cflat-brief,tinycfa-brief,caulfield2023acfa-brief,lofat-brief,ammar2024bridging-brief,yadav2023whole-brief,zhang2021recfa-brief,neto2023isc-brief,atrium-brief,debes2023zekra-brief,wang2023ari-brief,litehax-brief,dialed-brief,abera2019diat-brief,kuang2020ra-brief} or targeted higher-end systems 
  \cite{davi2015hafix-brief, christoulakis2016hcfi-brief, de2019fixer-brief,
zhou2020silhouette-brief, nyman2017cfi-brief, sullivan2016strategy-brief, almakhdhub2020mu-brief, das2016fine-brief, liljestrand2019pac-brief, 
gollapudi2023control-brief,park2022bratter-brief,yoo2022kernel-brief,murray2013sel4-brief,de2017sofia-brief,walls2019control-brief,fu2022fh-brief,peng2023cefi-brief}.
  
  \item A prototype implementation of \cacfi on {\sf openMSP430}, which includes 
  code instrumentation at compile-time and trusted software running in secure ROM.
  We also evaluate \cacfi performance, showing that it has a very low average 
  run-time overhead. All source code is publicly available at \cite{eilid-code}.
  \end{compactitem}

\section{Background \& Related Work}\label{sec:preliminary}
\subsection{Scope of Low-end Devices}\label{subsec:scope}
IoT devices vary greatly in computing power. For example, low-end devices (e.g., smart plugs or light bulbs) 
feature a single-core MCU and minimal amount of memory (e.g., a few KBs), while high-end devices 
(e.g., car infotainment systems) host multi-core MCUs and extensive memory (e.g., a few GBs).
In this work, we focus on the former.

A typical low-end MCU has an $8$- or $16$-bit Von Neumann architecture, 
running at $\leq$ $48$MHz with $\leq~64$KB of memory, e.g., TI MSP430 or AVR ATMega32.
SRAM is used as data memory (DMEM), typically ranging from $4$KB to $16$KB, while other
address space is available for program memory (PMEM).
Such devices usually run software atop ``bare metal'' with no memory management support, such as MMUs or MPUs.
Therefore, there is neither memory isolation nor privilege guarantees.

\subsection{Roots-of-Trust (RoTs) \& CASU} \label{subsec:rot}
Related work on \rot-s for low-end devices can be classified into: {\it passive} and {\it active} designs.
The former \cite{pistis-brief,sancus-brief,vrasedp-brief,rata-brief} detect malware presence on \prv 
via \ra. This involves \prv generating (upon a challenge from \vrf) a cryptographic
proof of its PMEM state and returning the result to \vrf, which decides/detects whether
\prv is compromised. Besides being passive, \ra imposes a non-negligible computational effort
on \prv.  Whereas, {\bf Active RoTs} \cite{casu-brief,garota-brief,lazarus-effect-brief,awdt-dominance-brief} 
continuously monitor \prv behavior to prevent or minimize the impact of compromise. 

\cacfi is built atop an active \rot, \textit{\textbf{CASU}} \cite{casu-brief} --
a hybrid hardware/software architecture that enforces 
run-time software immutability, while supporting authenticated software updates. It defends against code 
injection attacks (i.e., attacks that insert executable code by exploiting memory 
vulnerabilities \cite{return_into_libc}) by preventing unauthorized modifications of 
PMEM and any code execution from DMEM. The only means to modify PMEM is through secure updates.
To achieve this, \casu monitors several CPU hardware signals and triggers a reset if any violation 
is detected. \casu obviates the need for \ra between software updates, with minimal hardware modifications.

\subsection{Control-Flow Attacks}\label{subsec:cfa}
\begin{table*}[!hbtp]
\footnotesize
    \centering
    \resizebox{\textwidth}{!}
    {
    \begin{tblr}[
        note{$\dag$} = {forward-edge \cfi is achieved at a basic-block level}
    ]{
      colspec = {X[-1]X[1.8]X[-1]X[-1]X[-1]X[-1]X[2.5]X[7.3]},
      cells={halign=c,valign=m},
      vlines,
      hlines,
      stretch=0
    }
    {\bf Method} & {\bf Work} & {\bf RT}& {\bf F-edge} & {\bf B-edge} & {\bf Interrupt} & {\bf Platform} & {\bf Technique Summary} \\
    \SetCell[r=5]{c}{\cfi} &HAFIX\cite{davi2015hafix-brief} & \checkmark &  & \checkmark &  & Intel Siskiyou Peak & Extends Intel ISA with shadow stack \\    
    &HCFI\cite{christoulakis2016hcfi-brief} & \checkmark & \checkmark & \checkmark &  & Leon3 & Extends Sparc V8 ISA with shadow stack and labels \\
    &FIXER\cite{de2019fixer-brief} & \checkmark & \checkmark & \checkmark &  & RocketChip & Extends RISC-V ISA with shadow stack \\
    &Silhouette\cite{zhou2020silhouette-brief} & \checkmark & \checkmark & \checkmark & \checkmark & ARMv7-M & Uses ARM MPU for hardened shadow-stacks and labels \\    
    &CaRE\cite{nyman2017cfi-brief} & \checkmark & & \checkmark & & ARMv8-M & Uses ARM TrustZone for shadow stack \& nested interrupts \\
    \SetCell[r=4]{c}{\cfa} &Tiny-CFA\cite{tinycfa-brief} & & \checkmark & \checkmark & & openMSP430 & Hybrid \cfa with shadow stack \\
    &ACFA\cite{caulfield2023acfa-brief} & & \checkmark & \checkmark & \checkmark & openMSP430 & Active hybrid \cfa with secure auditing of code \\
    &LO-FAT\cite{lofat-brief} & & \checkmark & \checkmark & & Pulpino & Hardware-based \cfa solution \\
    &CFA+\cite{ammar2024bridging-brief} & & \checkmark & \checkmark & \checkmark & ARMv8.5-A & Leverages ARM's Branch Target Identification \\
    \SetCell[c=2]{c}{\boldmath ${\cacfi}$} & & \usym{2713} & \usym{2713} & \usym{2713} & \usym{2713} & {\bf opemMSP430} & {\bf Uses {\boldmath ${\casu}$} for shadow stack} \\
    \end{tblr}
    }
    \vspace*{0.1cm}
    \caption{\cfa and \cfi Techniques from Prior Work (RT: Real-time protection, F: Forward, B: Backward)} \label{table:cfi_technique}
    \vspace*{-0.3cm}
\end{table*}

Due to the lack of support for advanced memory in languages such as C/C++ and assembly 
(which are widely used in embedded systems software), devices are often vulnerable to 
control-flow attacks. There are two types of such attacks: code injection and code reuse.
The former inserts executable code by exploiting memory vulnerabilities in the user software,
e.g., stack overflow~\cite{return_into_libc}. These attacks can be thwarted by a simple memory 
protection policy, $W \oplus X$, preventing any execution from writable memory.
Code reuse attacks, on the other hand, deviate the software control flow 
to execute arbitrary (malicious) sequences of existing code. Relying on the target they aim to alter, 
control-flow attacks can be divided into forward- and backward-edge attacks.
The former manipulates forward edges in the Control-Flow Graph (CFG), such as indirect function calls and jumps,
while backward-edge attacks alter edges going backward, such as a return address.

There is a large body of research \cite{davi2015hafix-brief, christoulakis2016hcfi-brief, 
das2016fine-brief, de2019fixer-brief, sullivan2016strategy-brief, zhou2020silhouette-brief, 
almakhdhub2020mu-brief, nyman2017cfi-brief,bauer2022typro-brief,ge2017griffin-brief,
mashtizadeh2015ccfi-brief,tice2014enforcing-brief} guaranteeing \cfi on various embedded platforms.
Some \cite{davi2015hafix-brief,christoulakis2016hcfi-brief} extend an instruction set architecture with 
new \cfi instructions. Another \cfi approach \cite{zhou2020silhouette-brief,almakhdhub2020mu-brief,
nyman2017cfi-brief} relies on advanced features (e.g., MPUs, MMUs, and TEEs) to offer \cfi.
Nonetheless, since all prior techniques are geared for higher-end devices, 
they are not applicable to ``bare-metal'' devices that we target.

Another line of research \cite{cflat-brief,tinycfa-brief,caulfield2023acfa-brief,lofat-brief,
ammar2024bridging-brief,yadav2023whole-brief,zhang2021recfa-brief,neto2023isc-brief,atrium-brief,
debes2023zekra-brief,wang2023ari-brief,litehax-brief} focuses on \cfa using \ra.
\cfa records a log of the control-flow path taken by the software (on \prv).
Upon receiving \cfa request from \vrf, \prv generates an unforgeable integrity proof 
of this log and returns it to \vrf. Despite being computationally efficient for \prv 
(i.e., no validation/real-time protection), CFA only detects (does not prevent) 
control-flow deviations at run-time. It also incurs significant log 
storage and transmission costs. Moreover, log verification on \vrf is not trivial as it 
depends on the complexity of the code and the number of \prv-s.
As shown in Table \ref{table:cfi_technique}, \cacfi is the first \rot construction 
that offers real-time \cfi protection for low-end devices.

\begin{figure}[t]
  \centering  
  \captionsetup{justification=centering}  
  \includegraphics[width=0.95\columnwidth]{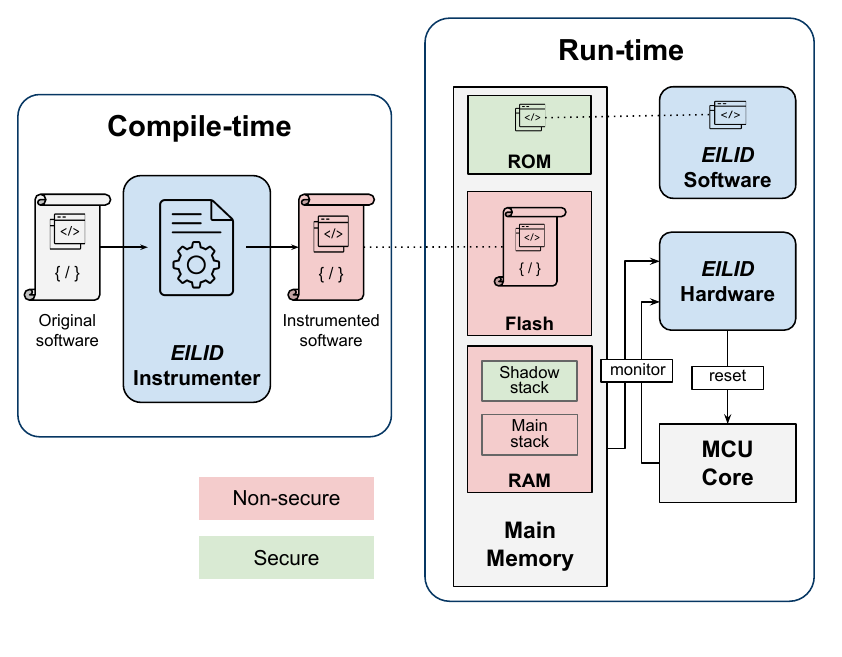}
  \vspace*{-.4cm}  
  \caption{\cacfi Design Overview}
  \label{fig:design}
  \vspace*{-0.35cm}
\end{figure}

\section{Overview \& Assumptions}
\subsection{\cacfi Overview}\label{subsec:overview}
As shown in Figure \ref{fig:design}, \cacfi has three components (colored blue): 
instrumenter \cacfiinst, trusted software \cacfisw, and hardware \cacfihw. 
We focus on the first two, since \cacfihw is based on \casu hardware 
and does not require any modification, except for the secure memory extension reserved for the shadow stack.
\cacfiinst instruments device software (which potentially has memory vulnerabilities) at 
compile-time to produce a \cfi-aware instrumented binary.
This binary is installed on \cacfi-enabled device, where \cacfihw continuously monitors  
software behavior and resets the device in case of \cfi violations.
At run-time, the software control flow is continuously validated by \cacfisw 
with the shadow stack support. Note that \cacfisw is immutable since it is housed in ROM.
Furthermore, the shadow stack is only accessible by \cacfisw, ensuring that 
control-flow metadata remains protected.

\subsection{Threat Model}\label{subsec:threat_model}
In line with other state-of-art \cfa/\cfi schemes 
\cite{bauer2022typro-brief,ge2017griffin-brief,mashtizadeh2015ccfi-brief,tice2014enforcing-brief}, 
we assume a powerful external adversary (\sadv) with comprehensive knowledge of all software 
running on the device, including all memory vulnerabilities, if any.
Since much of embedded software is written in memory-unsafe languages (e.g., C, C++, or Assembly),
we assume that it likely has (presumably unintentional) control-flow bugs.

\sadv can arbitrarily access any executable memory location at run-time.
It can also tamper with any data (e.g., return addresses, function pointers, and 
indirect function calls) on the stack and heap, thereby manipulating program control flow.
Moreover, \sadv can exploit vulnerabilities in Interrupt Service Routines (ISRs) to alter the program 
context (e.g., an ISR return address), thus forcing a deviation from the legal control flow.
As common in most related work, physical attacks and side-channels are out of scope.
Furthermore, \cacfi does not address non-control-data attacks. Protection against them 
is quite difficult and incurs prohibitive overhead for resource-constrained devices. 
Section \ref{sec:discussion} contains more information about this issue.

\subsection{Security Goals}\label{subsec:security_goal}
To enable \cacfi, the following security properties must hold.

\noindent{\bf [P1] Return Address Integrity:}
Function return addresses should be protected.
Any attempt to overwrite the return address, leading to a transition to an illegal 
(unintended) address, must be detected and handled properly.

\noindent{\bf [P2] Return-from-Interrupt Integrity:}
When an interrupt is triggered, the current program context is saved on the main stack, and the system 
branches to the appropriate ISR. After serving the interrupt, the stored context information is 
retrieved and execution is resumed where it left off. A memory vulnerability in an ISR allows 
modifications of the main stack where the context is kept, causing it to return to an illegal address.
Therefore, the interrupt context stored on the main stack must remain intact while the ISR runs.

\noindent{\bf [P3] Indirect Call Integrity:}
Unlike a direct branch which explicitly specifies the next instruction address,
an indirect call is invoked with the pointer containing the destination address.
Thus, that address can not be resolved at compile-time since it is unknown 
until the value is loaded from memory at run-time.

Besides these three properties, complete \cfi also requires integrity against indirect jumps.
However, indirect jumps are generally not critical because in practice they are only used for 
\verb+switch+ statements.  Jump-tables, generated to store indirect jump pointers for these 
statements, can be avoided by using compiler flags such as `\verb+-fno-jump-tables+' and 
`\verb+-fno-switch-tables+'. More details are in Section \ref{sec:discussion}.

\begin{table}
  \centering\captionsetup{justification = centering}  
  \resizebox{\columnwidth}{!} {
     \begin{tabular}{|Sc|Sc|Sc|Sc|Sc|} \hline
     & \multicolumn{4}{c|}{\bf Instructions} \\
     \hhline{~*4{-}}     
     & & & \bf {Return from} & \\     
     \multirow{-3}{*}{\bf Platform} & \multirow{-2}{*}{\bf Call} & \multirow{-2}{*}{\bf Return} & \bf {Interrupt} & \multirow{-2}{*}{\bf {Indirect Call}} \\ \hline
     \bf TI MSP430 & \texttt{CALL} & \texttt{RET} & \texttt{RETI} & \texttt{CALL} \\ \hline
     \bf AVR ATMega32 & \texttt{CALL} & \texttt{RET} & \texttt{RETI} & \texttt{RCALL, ICALL} \\ \hline
     \bf Microchip PIC16 & \texttt{CALL} & \texttt{RETURN} & \texttt{RETFIE} & \texttt{CALL, RCALL} \\ \hline
  \end{tabular}
  }
  \caption{Instruction Set in Low-end Platforms} 
  \label{table:inst_set}
  \vspace{-.4cm}
\end{table}

\section{\cacfi Design: \cacfiinst \& \cacfisw}\label{sec:design}

As stated in Section \ref{subsec:overview}, we use \casu hardware as \cacfihw without 
significant modifications. Consequently, we avoid introducing any new hardware overhead 
and preserve \casu's formally verified properties. Figure \ref{table:inst_set} shows
instruction sets that can be used by \cacfiinst on popular low-end MCU platforms.

\subsection{\cacfiinst}
\cacfiinst analyzes and instruments assembly code at compile-time.
For backward-edge \cfi, it discovers every function call statement (e.g., \verb+call+ in 
MSP430 and ATMega32) and adds a few instructions to jump to \cacfisw (\propa).
At run-time, before each function call site, the instrumented code resolves its return address 
(i.e., the next address of the function call site) and invokes \cacfisw to store the address on the shadow stack.

\cacfiinst also detects all function return instructions (e.g., \verb+RET+ in MSP430 and \verb+RETURN+ in PIC16).
It sequentially instruments the code to execute \cacfisw, verifying
that the return address in the current function context matches the one stored on the shadow stack.

Similarly, \cacfiinst instruments the binary at ISR prologues (entry points) to store interrupt context 
metadata (e.g., return address and status register) on the shadow stack (\propb). Meanwhile, at ISR epilogues, (i.e., right before 
return), instrumented code retrieves and verifies stored context meta against the current context.
\cacfiinst discovers ISR prologues by their reserved names, while
ISR epilogues are identified by return-from-interrupt instructions, e.g., \verb+RETI+ in MSP430 and \verb+RETFIE+ in PIC16.

Moreover, \cacfiinst enumerates entry points of all functions and puts them into a table for forward-edge \cfi (\propc).
\cacfiinst discovers indirect call instructions (e.g., \verb+CALL+ in MSP430 and \verb+RCALL, ICALL+ in ATMega32) and 
introduces a few extra instructions to verify the legitimacy of function addresses.
Note that \cacfi achieves function-level forward-edge \cfi because \sadv can redirect the indirect function call 
to another valid address in the function entry table. However, we believe that the chances of that are low due to 
the small number of functions in a typical low-end device software.

\subsection{\cacfisw}
At run-time, \cacfisw is invoked to store and validate function 
or ISR return addresses provided by \cacfiinst as function arguments (\propa \& \propb).
Before executing an indirect function call, \cacfisw is triggered to verify the legitimacy 
of the function by searching for it in the table (\propc).
If any \cfi validations fail, \cacfihw resets the device, thus thwarting control-flow attacks.

\begin{table}[!hbtp]
\footnotesize
	\centering	
    \begin{tabular}{|m{1.1cm}||m{6.5cm}|}
        \hline
        \multicolumn{1}{|c||}{\bf Registers} & \multicolumn{1}{c|} {\bf Description} \\
        \hline
        \multicolumn{1}{|c||}{{r4}} & Used as an argument of \verb+S_EILID_init()+ \\
        \hline
        \multicolumn{1}{|c||}{r5} & Used as a pointer to the shadow stack's current index \\
        \hline
        \multicolumn{1}{|c||}{r6, r7} & Used as an argument of other \verb+S_EILID+ functions \\
        \hline
	\end{tabular}
	\caption{Reserved Registers for \cacfi}
	\label{table:reserved_registers}
	\vspace*{-0.2cm}
\end{table}

\section{Implementation}\label{sec:implementation}
\cacfi is implemented atop openMSP430 \cite{openmsp430-brief}, an open-sourced 16-bit MCU core. 
The code is synthesized using Xilinx Vivado 2023.1, and
the synthesized design is then deployed on a Basys3 Artix-7 FPGA board for prototyping and evaluation.
All \cacfi source code is open-sourced at \cite{eilid-code}.

\cacfiinst is a Python script with $\approx~200$ lines of code.
To facilitate instrumentation, general-purpose user registers, \verb+r4+-\verb+r7+
are reserved for \cacfi, which are seldom used.
Their usage is summarized in Table \ref{table:reserved_registers}.
If any of those registers are in use by device software, 
merely two instructions are additionally needed: (1) push the register value on the main stack before its usage, e.g., \verb+push r4+,
and (2) pop the value from the stack after it is used, e.g., \verb+pop r4+.

We allocate $256$ bytes of secure DMEM (realized by \cacfi hardware) for the shadow stack, exclusive to \cacfisw.
It can store $\leq~128$ return addresses and the interrupt context.
Since MSP430 devices usually have tens of KB of memory, reserving  $256$ bytes for \cacfi should be acceptable.
Also, the memory region that stores the function return address is freed up and ready for re-use 
once the current function returns. Therefore, we believe it can accommodate control-flow metadata of all 
tasks running on typical low-end commodity devices.
Nevertheless, the shadow stack size is configurable based on memory constraints and software complexity.

\subsection{\cacfiinst}\label{subsec:impl_inst}

\begin{figure}[t]
  \centering
  \captionsetup{justification=centering}
  \includegraphics[width=0.9\columnwidth]{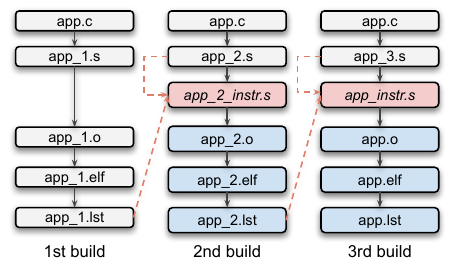}
  \caption{\cacfi Instrumented Compilation}
  \label{fig:compile}
  \vspace{0cm}
\end{figure}

\lstdefinelanguage
   [x64]{Assembler}     % add a "x64" dialect of Assembler
   [x86masm]{Assembler} % based on the "x86masm" dialect
   %
   % with these extra keywords:
   {morekeywords={mov.b,jn,jlo,cmp.b}}
   
\lstset{language=[x64]{Assembler},
	basicstyle={\tiny\ttfamily},
	showstringspaces=false,
	frame=single,
	xleftmargin=2em,
	framexleftmargin=3em,
	numbers=left, 
	numberstyle=\tiny,
	commentstyle={\tiny\itshape},
	keywordstyle={\tiny\ttfamily\bfseries},
	keywordstyle=\color{blue}\tiny\ttfamily\ttfamily,
	stringstyle=\color{red}\tiny\ttfamily,
        commentstyle=\color{black}\tiny\ttfamily,
        morecomment=[l][\color{magenta}]{\%},
        breaklines=true
}

\begin{figure}
    \centering
    \begin{minipage}{0.49\linewidth}
\begin{lstlisting}[xleftmargin=.18\textwidth, xrightmargin=.07\textwidth]


call #foo
\end{lstlisting}
        \centering
        \scriptsize{(a) Original}
    \end{minipage}
    \begin{minipage}{0.49\linewidth}
\begin{lstlisting}[xleftmargin=.18\textwidth, xrightmargin=.07\textwidth]
mov #e200, r6
call #NS_EILID_store_ra
call #foo
\end{lstlisting}
        \centering
        \scriptsize{(b) Instrumented}
    \end{minipage}
    \caption{Instrumentation before Function Call}
    \label{fig:instr_1}
    \begin{minipage}{0.49\linewidth}
\begin{lstlisting}[xleftmargin=.18\textwidth, xrightmargin=.07\textwidth]
foo:
    ...
    

    ret
\end{lstlisting}
        \centering
        \scriptsize{(a) Original}
    \end{minipage}
    \begin{minipage}{0.49\linewidth}
\begin{lstlisting}[xleftmargin=.18\textwidth, xrightmargin=.07\textwidth]
foo:
    ...
    mov @(r1), r6
    call #NS_EILID_check_ra
    ret
\end{lstlisting}
        \centering
        \scriptsize{(b) Instrumented}
    \end{minipage}
    \caption{Instrumentation before Function Return}
    \vspace*{-.2cm}
    \label{fig:instr_2}
\end{figure}

\begin{figure}
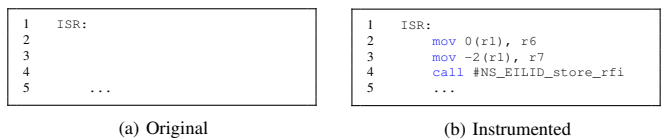

	\centering
	\begin{minipage}{0.49\linewidth}
\begin{lstlisting}[xleftmargin=.18\textwidth, xrightmargin=.07\textwidth]
ISR:



    ...
\end{lstlisting}
            \centering
            \scriptsize{(a) Original}
        \end{minipage}
        \begin{minipage}{0.49\linewidth}
\begin{lstlisting}[xleftmargin=.18\textwidth, xrightmargin=.07\textwidth]
ISR:
    mov 0(r1), r6
    mov -2(r1), r7
    call #NS_EILID_store_rfi
    ...
\end{lstlisting}
            \centering
	    \scriptsize{(b) Instrumented}
        \end{minipage}
	\caption{Instrumentation at ISR Entry point}
         \vspace*{-.4cm}
	\label{fig:instr_3}
\end{figure}

\begin{figure}
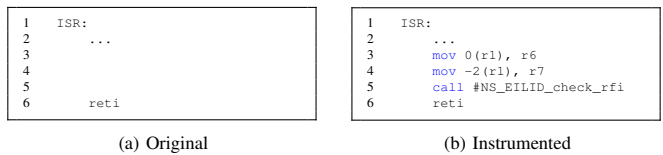

	\centering
	\begin{minipage}{0.49\linewidth}
\begin{lstlisting}[xleftmargin=.18\textwidth, xrightmargin=.07\textwidth]
ISR:
    ...


    
    reti
\end{lstlisting}
            \centering
            \scriptsize{(a) Original}
        \end{minipage}
        \begin{minipage}{0.49\linewidth}
\begin{lstlisting}[xleftmargin=.18\textwidth, xrightmargin=.07\textwidth]
ISR:
    ...
    mov 0(r1), r6
    mov -2(r1), r7
    call #NS_EILID_check_rfi
    reti
\end{lstlisting}
            \centering
	    \scriptsize{(b) Instrumented}
        \end{minipage}
	\caption{Instrumentation before ISR Return}
	\label{fig:instr_4} 
    \vspace*{-.2cm}
\end{figure}

\begin{figure}
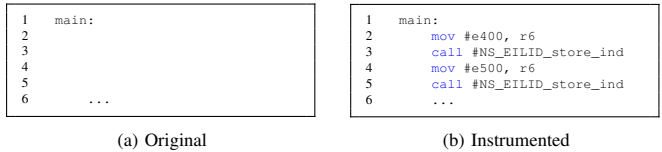

	\centering
	\begin{minipage}{0.49\linewidth}
\begin{lstlisting}[xleftmargin=.18\textwidth, xrightmargin=.07\textwidth]
main:




    ...
\end{lstlisting}
            \centering
            \scriptsize{(a) Original}
        \end{minipage}
        \begin{minipage}{0.49\linewidth}
\begin{lstlisting}[xleftmargin=.18\textwidth, xrightmargin=.07\textwidth]
main:
    mov #e400, r6
    call #NS_EILID_store_ind
    mov #e500, r6
    call #NS_EILID_store_ind
    ...
\end{lstlisting}
            \centering
	    \scriptsize{(b) Instrumented}
        \end{minipage}
	\caption{Instrumentation at Main Function Entry point}
    \vspace{-.3cm}
	\label{fig:instr_5}
\end{figure}

\begin{figure}
	\centering
	\begin{minipage}{0.49\linewidth}
\begin{lstlisting}[xleftmargin=.18\textwidth, xrightmargin=.07\textwidth]


call r13
\end{lstlisting}
            \centering
            \scriptsize{(a) Original}
        \end{minipage}
        \begin{minipage}{0.49\linewidth}
\begin{lstlisting}[xleftmargin=.18\textwidth, xrightmargin=.07\textwidth]
mov r13, r6
call #NS_EILID_check_ind
call r13
\end{lstlisting}
            \centering
	    \scriptsize{(b) Instrumented}
        \end{minipage}
	\caption{Instrumentation before Indirect Function Call}
     \vspace*{-.4cm}
	\label{fig:instr_6} 
\end{figure}

\begin{figure}
	\centering
	\subfigure[\cacfi Software Flow \label{fig:sw_flow}]
	{\includegraphics[width=0.6\columnwidth]{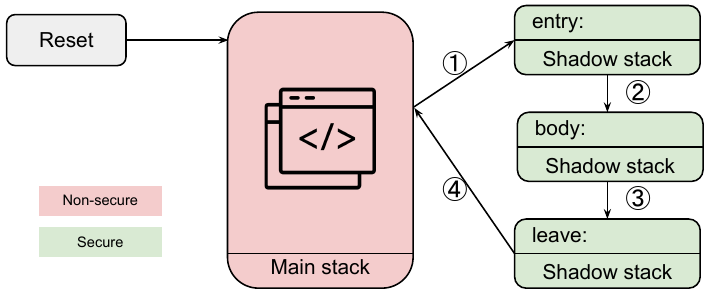}}
    \hspace{.3em}
	\subfigure[\centering Shadow Stack Example \label{fig:shadow_stack}]
	{\includegraphics[width=0.35\columnwidth]{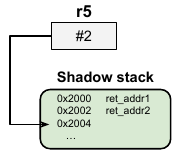}}
      \vspace*{-0.3cm}
	\caption{\cacfi Software Implementation}
	\label{fig:cacfi_sw} 
\end{figure}

Figure \ref{fig:compile} shows the iterated compile process with \cacfiinst.
It takes as input two files: (1) \verb+*.lst+ to discover function return addresses, 
and (2) \verb+*.s+ to be instrumented.
Its output is \verb+*instr.s+ file, which is colored red in the iterations.
Three iterations are needed due to the input file \verb+*.lst+, generated after 
the build is completed, and shifted addresses in the second loop after the first-iteration instrumentation.
This adjustment is essential because the return address of each function changes 
from the instructions introduced during the second iteration.
For example, if the return address of \verb+foo()+ is initially  0x$200$ (\verb+app_2.s+), 
it could be modified to 0x$210$ if $8$ instructions are added ahead of it in 
\verb+app_2_instr.s+ file. This also results in increased file sizes (colored red and blue).
Because \verb+app_2.lst+ already shifted addresses, it is directly used as input of \cacfiinst
at the last loop with no further changes.

Figures \ref{fig:instr_1} and \ref{fig:instr_2} show the instrumentation needed to achieve \propa.
Before jumping to the function, its next instruction address (0xe$200$,
the return address of \verb+foo()+) is loaded to \verb+r6+ as an argument of \verb+NS_EILID+
\verb+_store_ra()+. For example, if the function call address is 0x$100$, its return address would be 
0x$102$ or 0x$104$, depending on its instruction size.
Then, it invokes \verb+NS_EILID_store_ra()+ to store this address on the shadow stack.
Because the function return address was pushed on the main stack (\verb+r1+ in MSP430) at the function prologue,
this value is loaded to \verb+r6+ and \verb+NS_EILID_check_ra()+ is called to check before \verb+ret+ instruction.

Figures \ref{fig:instr_3} and \ref{fig:instr_4} show how the code is instrumented to 
ensure \propb. In MSP430 architecture, when an interrupt is triggered, the current instruction address and the 
status register (\verb+r2+) are pushed on the main stack. We consider these two values as the interrupt context.
At the ISR entry point, the interrupt context is loaded to \verb+r6+ and \verb+r7+ as arguments, and 
in turn, \verb+NS_EILID_store_rai()+ is called. Before its return, the system loads the interrupt context 
to \verb+r6+ and \verb+r7+, and validates the context in \verb+NS_EILID_check_rai()+.

As illustrated in Figure \ref{fig:instr_5}, at the beginning of \verb+main()+ function (once the device boot sequence 
completes), a few instructions are introduced calling \verb+NS_EILID_store_+ \verb+ind_func()+. The code stores all 
legitimate function addresses of the device software in a table.
Figure \ref{fig:instr_6} shows that the target address of an indirect function is checked via 
\verb+NS_EILID_check_ind_func()+. It makes sure that
the address is present in the table before each indirect function call.

\subsection{\cacfisw}\label{subsec:impl_sw}
\cacfisw is composed of three sections: entry, body, and leave.
The entry section is the only legal entry point that can switch to a secure state, thus 
minimizing the attack surface. The body section houses all \verb+S_EILID+ functions.
Before transitioning to execute non-secure software, the exit section must be passed through.
\cacfihw (i.e., \casu hardware) ensures atomicity of all authorized functions running in 
the secure state by resetting the device if any violation occurs.

Figure \ref{fig:sw_flow} illustrates the software flow on \cacfi-enabled devices.
After a reset, the device runs its normal software.
Instrumented code jumps to \cacfisw, switching to the secure state -- \ding{172}.
Within the entry section, \verb+r4+ determines which \verb+S_EILID+ function is invoked.
For example, if \verb+r4+$==1$, it branches to \verb+S_EILID_store_ra()+ in the body section -- \ding{173}.
In the function, the return address is stored on the shadow stack.
Finally, it branches to \verb+S_EILID_exit()+ in the exit section -- \ding{174},
and resumes normal device software -- \ding{175}.

Recall that \verb+r5+ is reserved for the shadow stack index. For example, if \verb+r5+$==2$, 
as shown in Figure \ref{fig:shadow_stack}, the next return address is stored at 
0x$2000+2*(2-1)=$ 0x$2002$ (each memory size is 2 bytes in 16-bit architecture).
When \verb+S_EILID_store_ra()+ is invoked, it stores the address at 
0x$2000+2*$\verb+r5+$=$0x$2004$ and increments \verb+r5+ by $1$.
This obviates the need for memory access to the shadow stack to maintain its index,
thus improving performance. When \verb+S_EILID_check_ra()+ is executed,
the stored address on the shadow stack is popped off and \verb+r5+ is decremented by $1$.

\section{Evaluation}\label{sec:evaluation}
\begin{figure}	
    \centering
    \subfigure[\cacfi LUT overhead \label{fig:hw_lut}]
    {\includegraphics[width=0.49\columnwidth]{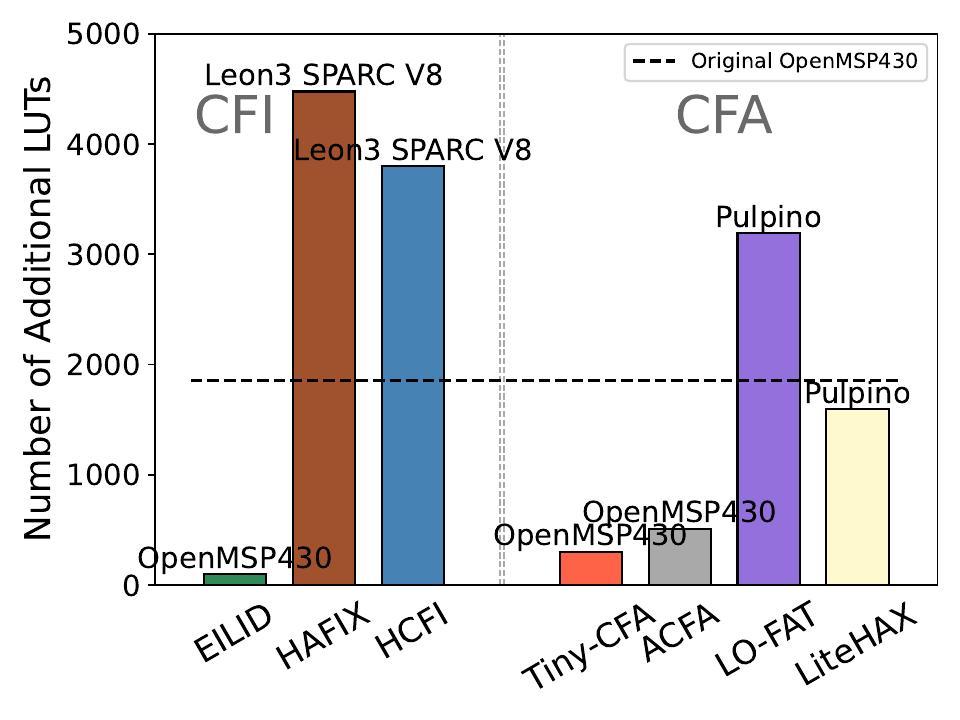}}
    \subfigure[\cacfi Register overhead \label{fig:hw_reg}]
    {\includegraphics[width=0.49\columnwidth]{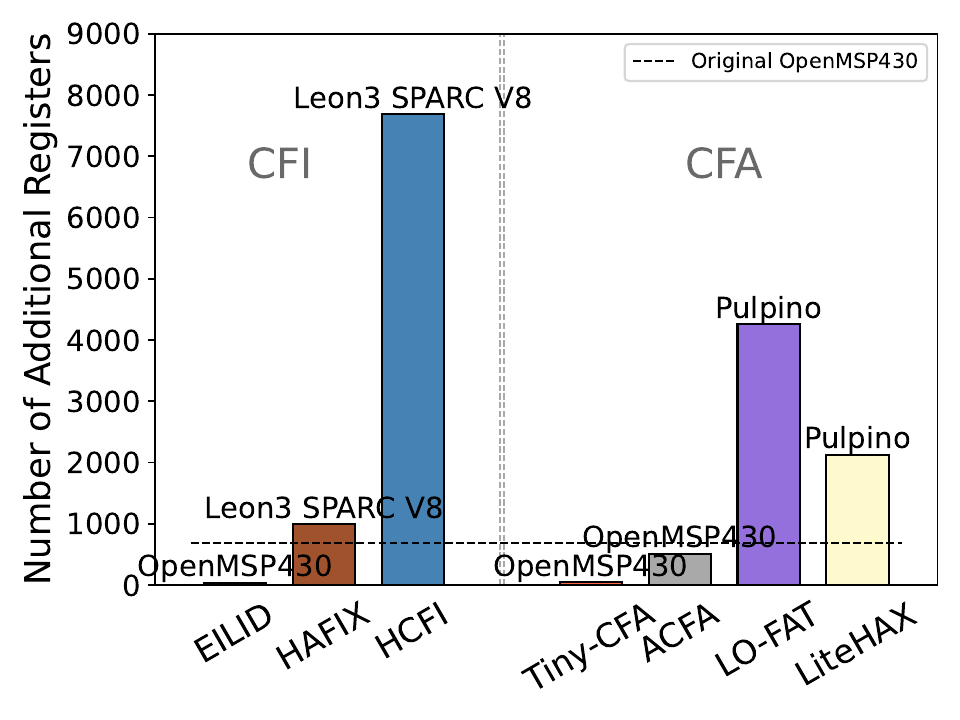}}
    \caption{\centering \cacfi Hardware Overhead Comparison}	
    \label{fig:cacfi_hw_overhead} 
\end{figure}

\noindent {\bf \cacfi Hardware Overhead:}
Recall that \cacfi hardware overhead is entirely derived from \casu hardware.
Figure \ref{fig:cacfi_hw_overhead} compares hardware costs to prior \cfi 
techniques (HAFIX \cite{davi2015hafix-brief} and HCFI \cite{christoulakis2016hcfi-brief}) 
as well as \cfa schemes (Tiny-CFA \cite{tinycfa-brief}, ACFA \cite{caulfield2023acfa-brief}, 
LO-FAT \cite{lofat-brief}, and LiteHAX \cite{litehax-brief}).
Since most of these techniques are implemented on different architectures and platforms, 
it is difficult to compare them directly. However, we believe that \cacfi hardware cost is minimal 
because: (i) Tiny-CFA and ACFA, implemented on the same platform (openMSP430),
have substantially higher hardware overhead, and (ii) other techniques geared for 32-bit architectures 
have more computing resources, as shown in Figure \ref{fig:cacfi_hw_overhead}.
In absolute numbers, \cacfi augments Look-Up Tables (LUTs) by $99$ (5.3\%) and registers -- 
by $34$ (4.9\%), over the openMSP430 baseline. Whereas, Tiny-CFA and ACFA increase: (1) LUTs by $302$ (16.2\%) 
and $501$ (26.9\%), respectively, and (2) registers by $44$ (6.4\%) and $946$ (136.7\%), respectively.
Furthermore, LO-FAT and LiteHAX require $216$KB and $158$KB of RAM, respectively \cite{apex-brief}.
Such sizes far exceed the entire addressable memory ($64$KB) of a 16-bit CPU of MSP430.
This supports our claim that low-end devices can not use prior techniques designed for higher-end platforms.

\begin{table}[!btp]
    \centering
    \resizebox{\columnwidth}{!} {
    \begin{tabular}{ V{2}cV{2}cV{2}cV{2}cV{2}cV{2}cV{2} }
        \hlineB{2}
        \multicolumn{2}{V{2}cV{2}}{\bf Software} & {\bf Compile-time} & {\bf Binary size} & {\bf Running time} \\
        \hlineB{2}
        \multirow{3}{4em} {\centering \bf Light Sensor} 
            & Original   & $321ms$                & $233$ byte       & $251{\mu}s$ \\
        \cline{2-5}
            & \cacfi     & $419ms$                & $246$ byte       & $277{\mu}s$ \\
        \cline{2-5}
            & {\centering \bf diff} & $98ms$ {\bf (30.53\%)} & $13$ byte {\bf (5.58\%)}  & $26{\mu}s {\bf (10.36\%)}$ \\            
        \hlineB{2}
        \multirow{3}{5em} {\centering \bf Ultrasonic Ranger} 
            & Original   & $334ms$                & $296$ byte       & $2,094{\mu}s$ \\
        \cline{2-5}
            & \cacfi     & $423ms$                & $349$ byte       & $2,303{\mu}s$ \\
        \cline{2-5}
            & {\centering \bf diff} & $89ms$ {\bf (26.65\%)} & $53$ byte {\bf (17.91\%)} & $209{\mu}s$ {\bf (9.98\%)}\\
        \hlineB{2}
        \multirow{3}{4em} {\centering \bf Fire Sensor} 
            & Original   & $341ms$                & 465 byte       & $4,105{\mu}s$ \\
        \cline{2-5}
            & \cacfi     & $484ms$                & 565 byte       & $4,648{\mu}s$ \\
        \cline{2-5}
            & {\centering \bf diff} & 143ms {\bf (41.94\%)} & 100 byte {\bf (21.51\%)} & $543 {\mu}s$ {\bf (13.23\%)}\\
        \hlineB{2}
        \multirow{3}{4em} {\centering \bf Syringe Pump} 
            & Original   & $318ms$                & 274 byte       & $2,151{\mu}s$ \\
        \cline{2-5}
            & \cacfi     & $458ms$                & 308 byte       & $2,265{\mu}s$ \\
        \cline{2-5}
            & {\centering \bf diff} & 140ms {\bf (44.03\%)} & 34 byte {\bf (12.41\%)} & $114 {\mu}s$ {\bf (5.30\%)}\\
        \hlineB{2}
        \multirow{3}{4em} {\centering \bf Temp Sensor} 
            & Original   & $351ms$                & 305 byte       & $1,257{\mu}s$ \\
        \cline{2-5}
            & \cacfi     & $465ms$                & 325 byte       & $1,327{\mu}s$ \\
        \cline{2-5}
            & {\centering \bf diff} & 114ms {\bf (32.48\%)} & 20 byte {\bf (6.56\%)} & $70 {\mu}s$ {\bf (5.57\%)}\\
        \hlineB{2}
        \multirow{3}{4em} {\centering \bf Charlie-plexing} 
            & Original   & $360ms$                & 325 byte       & $4,930{\mu}s$ \\
        \cline{2-5}
            & \cacfi     & $455ms$                & 342 byte       & $5,146{\mu}s$ \\
        \cline{2-5}
            & {\centering \bf diff} & 95ms {\bf (26.39\%)} & 17 byte {\bf (5.23\%)} & $216 {\mu}s$ {\bf (4.38\%)}\\
        \hlineB{2}
        \multirow{3}{4em} {\centering \bf Lcd Sensor} 
            & Original   & $370ms$                & 604 byte       & $4,877{\mu}s$ \\
        \cline{2-5}
            & \cacfi     & $474ms$                & 642 byte       & $5,005{\mu}s$ \\
        \cline{2-5}
            & {\centering \bf diff} & 104ms {\bf (38.11\%)} & 38 byte {\bf (6.29\%)} & $128 {\mu}s$ {\bf (2.62\%)}\\
        \hlineB{2}
        \multicolumn{2}{V{2}cV{2}} {\centering \bf Average Overhead} & {\bf 34.30\%} & {\bf 10.78\%} & {\bf 7.35\%} \\
        \hlineB{2}
    \end{tabular}
    }
    \vspace{.1cm}
    \caption{\cacfi Software overhead} \label{table:sw_overhead}
\end{table}

\noindent {\bf \cacfi Software Overhead: }
Note that there are no benchmarking tools available for low-end devices, i.e., 8-bit or 16-bit architectures.
Thus, we instead used several publicly available practical applications to evaluate \cacfi software overhead:  
[\verb+LightSensor+, \verb+FireSensor+, \verb+UltrasonicRanger+]%
\footnote{\url{https://github.com/Seeed-Studio/LaunchPad_Kit}},
\verb+SyringePump+%
\footnote{\url{https://github.com/manimino/OpenSyringePump}}, and
[\verb+Temp+ \verb+Sensor+, \verb+Charlieplexing+, \verb+LcdSensor+]%
\footnote{\url{https://github.com/ticepd/msp430-examples}}.
Some of these applications, ported to run on openMSP430, have been used to motivate and evaluate prior 
\cfa and \cfi techniques \cite{cflat-brief,tinycfa-brief,dialed-brief,caulfield2023acfa-brief}.

Table \ref{table:sw_overhead} presents experimental results for these applications at 
compile- and run-time. Applications are built on an Ubuntu 20.04 LTS 
desktop with an Intel i5-11400 processor running at $2.6$GHz with $16$GB of RAM.
Compile-time in each scenario is measured on average over $50$ iterations.
In each case, compile-time increases by $\leq~44.03\%$, which is reasonable 
considering that \cacfi requires three compile iterations (see Figure \ref{fig:compile}).
The binary size grows by $\leq~21.51\%$ across all applications.

Run-time overhead, measured using Vivado 2022.1 behavioral simulation running at 100MHz, 
ranges from $2.62\%$ to $13.23\%$. Given infrequent use of indirect function calls in
typical low-end device software, run-time overhead primarily arises from ensuring function 
return address (\propa) and return-from-interrupt (\propb) integrity.
The instrumentation overhead per function call or interrupt is $\approx~25.2{\mu}s$.
In more detail, storing control-flow values takes $\approx~11.8{\mu}s$,
while checking these values takes $\approx~13.4{\mu}s$.
The number of instructions introduced for storing and checking is $26$ and $29$, 
respectively. We believe these results are reasonable, considering  
the context switch overhead between normal software and \cacfisw.

\section{Discussion} \label{sec:discussion}
\noindent{\bf Non-control-data Attacks:}
Non-control-data attacks \cite{chen2005non-brief} do not directly corrupt control-flow metadata.
Instead, these attacks manipulate application data, involving environmental data (i.e., data of sensor/actuator), 
user inputs, and loop/conditional variables. For example, \sadv can corrupt the heap memory to modify a 
loop variable in a \verb+if+ statement, thus skipping an integrity check before running a new task.
Since such actions do not alter control-flow metadata, they can circumvent various control-flow 
protections that only monitor that metadata. 
To ensure integrity of such data, all memory used for non-control data must be monitored and protected.

However, monitoring all non-control data incurs an excessive software overhead. Consider a simple loop, \verb|for (i=0; i<100; i++)|.
Since manipulation of \verb+i+ influences the execution sequence, it should be protected to maintain
data-flow integrity. However, preserving the value of `\verb+i+' would require storing it $100$
times and verifying -- $99$ times, since the initial value of $0$ is stored without checking.
Also, tracing such data outside the current function scope requires additional processing, when 
it is passed as function arguments or return values.
This results in a prohibitive memory overhead that low-end devices can not afford.

\noindent{\bf Indirect Jumps:}
Consistent with prior work \cite{lu2019does-brief,ge2016fine-brief,amit2019jumpswitches-brief} and as mentioned earlier,
\cacfi does not consider indirect jumps.
They are primarily used to implement \verb+switch+ statements, which are converted into jump-tables at compilation. 
These jump-tables are read-only and boundary-checked at run-time to ensure that
they do not leave the function boundary. \cacfiinst avoids generating jump instructions by 
compiling with flags, such as \verb+-fno-jump-tables+ and \verb+-fno-switch-tables+.
Also, a compile-time warning is raised if there are indirect jumps besides \verb+switch+ statements.

\noindent{\bf Recursion:}
\cacfi does not handle recursion because of its excessive memory overhead. Specifically, 
recursion consumes significant stack space and hence it is rarely used in embedded systems software.
A function call in MSP430 requires at least $2$ bytes for the return address, plus a few more bytes to 
store local variables. For example, on a device with only $2$KB of RAM, a recursive function can be invoked at 
most $204$ times with $2$ integer variables ($8$ bytes) in recursion.
This number is significant if other functions or global variables are present.
Thus, it is generally unadvisable to use recursion on low-end devices;
instead, it makes more sense to convert them into iterative statements using loops.

\section{Conclusions}\label{sec:conclusion}
This paper constructed \cacfi, a prevention-based \rot architecture that assures software integrity 
and prevents run-time attacks on low-end devices. Its evaluation on openMSP430 shows 
that \cacfi achieves a low average run-time overhead of $\approx~7.5\%$ across seven real-world 
applications.

\section*{Acknowledgment}
We thank DATE 2025 reviewers for their constructive feedback. This work was supported in part by 
funding from NSF Award SATC-1956393, NSA Awards H98230-20-1-0345 and H98230-22-1-0308, 
as well as a DARPA subcontract from Peraton Labs.

\balance
\bibliographystyle{ieeetr}
\bibliography{references}

\end{document}